\documentclass[twocolumn,showpacs,preprintnumbers,amsmath,amssymb]{revtex4}
\usepackage{graphicx}
\usepackage{amssymb,amsmath}
\usepackage{bm,slashed}
\usepackage{dcolumn}

\begin{document}

\preprint{}

\title{No Drama Quantum Electrodynamics? }

\author{A. Akhmeteli}
\email{akhmeteli@ltasolid.com}
\affiliation{%
LTASolid Inc.\\
10616 Meadowglen Ln. 2708\\
Houston, TX 77042, USA}%


\homepage{http://www.akhmeteli.org}

\date{\today}

\begin{abstract}
This article builds on recent work (A. Akhmeteli, Int'l Journ. of Quantum Information,  vol. 9, Suppl. (2011) p. 17, and A. Akhmeteli, Journ. Math. Phys., vol. 52 (2011) p. 082303), providing a theory that is based on spinor electrodynamics, is described by a system of partial differential equations in 3+1 dimensions, but reproduces unitary evolution of a quantum field theory in the Fock space. To this end, after introduction of a complex four-potential of electromagnetic field, which generates the same electromagnetic fields as the initial real four-potential, spinor field is algebraically eliminated from the equations of spinor electrodynamics. It is proven that the resulting equations for electromagnetic field describe independent evolution of the latter and can be embedded into a quantum field theory using a generalized Carleman linearization procedure. The theory provides a simple  and at least reasonably realistic model, valuable for interpretation of quantum theory. The issues related to the Bell theorem are discussed.
\end{abstract}


\maketitle

\section{Introduction}
\label{intro}
In an earlier article (Ref.~\cite{Akhmeteli-IJQI}), this author discussed a possibility of a ``no drama'' quantum theory, as simple (in principle) as classical electrodynamics --- a local realistic theory described by a system of partial differential equations in 3+1 dimensions, but reproducing unitary evolution of a quantum field theory in the Fock space. In particular, it was shown there that the matter field can be algebraically eliminated from the equations of scalar electrodynamics (the Klein-Gordon-Maxwell electrodynamics) in the unitary gauge, the resulting equations for electromagnetic field describe independent evolution of the latter and can be embedded into a quantum field theory. The issue of the Bell theorem was discussed in detail using arguments of nightlight and E. Santos. The analysis can be summarized as follows. While the Bell inequalities cannot be violated in the theories of Ref.~\cite{Akhmeteli-IJQI}, there are some reasons to believe these inequalities cannot be violated either in experiments or in quantum theory: on the one hand,
there is no loophole-free experimental evidence of violations of the Bell inequalities (see, e.g., Ref.~\cite{Aspel}), on the other hand, to prove that the inequalities can be violated in quantum theory, one needs to use the theory of quantum measurements, e.g., the projection postulate. However, such postulate is in contradiction with the standard unitary evolution (this is the well-known problem of measurement in quantum theory), as such postulate introduces irreversibility and turns a superposition of states into their mixture. Therefore, mutually contradictory assumptions are required to prove the Bell theorem (not in the part related to the derivation of the Bell inequalities for local realistic theories, but in the part related to violations of the Bell inequalities in quantum theory), so it can be circumvented if the projection postulate is rejected in favor of unitary evolution. The reader is referred to Ref.~\cite{Akhmeteli-IJQI} for the detailed analysis and the references.

The extension of the above results to spinor electrodynamics (the Dirac-Maxwell electrodynamics) offered in Ref.~\cite{Akhmeteli-IJQI} was much more limited and less satisfactory, as, instead of the Dirac equation, its modification for a limited class of functions was used. The root of the problem was that, while a scalar field can always be made real (at least locally) by a gauge transform (Ref.~\cite{Schroed}), this is not true for a spinor field described by the Dirac equation. Recently, however, this line of research produced a most important spin-off (Ref.~\cite{Akhmeteli-JMP}): it was shown that, in a general case, three out of four complex components of the Dirac spinor can be algebraically eliminated from the Dirac equation in electromagnetic field, and the remaining component can be made real (at least locally) by a gauge transform. Thus, on the one hand, the Dirac equation is generally equivalent to a fourth order partial differential equation for just one real component, on the other hand, most results of Ref.~\cite{Schroed} for scalar fields and scalar electrodynamics were extended to spinor fields and spinor electrodynamics. This opened a way to the main result of the present article (announced in Ref.~\cite{Akhmeteli-JPHC}): a much more satisfactory extension of the results of Ref.~\cite{Akhmeteli-IJQI} to spinor electrodynamics, which is more realistic than scalar electrodynamics.

\section{Scalar electrodynamics}
\label{sec:1}
Let us first consider scalar electrodynamics, both to illustrate the general approach using this simpler case and to present a proof for scalar electrodynamics that is significantly improved compared to Ref.~\cite{Akhmeteli-IJQI}, as it does not contain the nasty square roots (see, e.g., equation (15) of Ref.~\cite{Akhmeteli-IJQI}).

The equations of scalar electrodynamics are as follows:
\begin{equation}\label{eq:pr7}
(\partial^\mu+ieA^\mu)(\partial_\mu+ieA_\mu)\psi+m^2\psi=0,
\end{equation}
\begin{equation}\label{eq:pr8}
\Box A_\mu-A^\nu_{,\nu\mu}=j_\mu,
\end{equation}
\begin{equation}\label{eq:pr9}
j_\mu=ie(\psi^*\psi_{,\mu}-\psi^*_{,\mu}\psi)-2e^2 A_\mu\psi^*\psi.
\end{equation}
The metric tensor used to raise and lower indices is (Ref.~\cite{Itzykson})
\begin{equation}\label{eq:d1ps}
\nonumber
g_{\mu\nu}=g^{\mu\nu}=\left( \begin{array}{cccc}
1 & 0 & 0 & 0\\
0 & -1 & 0 & 0\\
0 & 0 & -1 & 0\\
0 & 0 & 0 & -1 \end{array} \right), g_\mu^\nu=\delta_\mu^\nu.
\end{equation}

The complex charged matter field $\psi$ in scalar electrodynamics (equations (\ref{eq:pr7}), (\ref{eq:pr8}), and (\ref{eq:pr9})) can be made real by a gauge transform (at least locally), and the equations of motion in the relevant gauge (unitary gauge) for the transformed four-potential of electromagnetic field $B^{\mu}$ and real matter field $\varphi$ are as follows (Ref.~\cite{Schroed}):
\begin{equation}\label{eq:pr10}
\Box\varphi-(e^2 B^\mu B_\mu-m^2)\varphi=0,
\end{equation}
\begin{equation}\label{eq:pr11}
\Box B_\mu-B^\nu_{,\nu\mu}=j_\mu,
\end{equation}
\begin{equation}\label{eq:pr12}
j_\mu=-2e^2 B_\mu\varphi^2.
\end{equation}

The following unexpected result was proven in Ref.~\cite{Akhmeteli-IJQI}: the equations obtained from equations ~(\ref{eq:pr10}), (\ref{eq:pr11}), and (\ref{eq:pr12}) after natural elimination of the matter field form a closed system of partial differential equations and thus describe independent dynamics of electromagnetic field. The detailed wording is as follows: if components of the four-potential of the electromagnetic field and their first derivatives with respect to time are known in the entire space at some time point, the values of their second derivatives with respect to time can be calculated for the same time point, so the Cauchy problem can be posed, and integration yields the four-potential in the entire space-time.

To eliminate the matter field $\varphi$ from equations ~(\ref{eq:pr10}), (\ref{eq:pr11}), and (\ref{eq:pr12}), let us use a substitution $\Phi=\varphi^2$ first. For example, as
\begin{equation}\label{eq:pr1q}
\Phi_{,\mu}=2\varphi\varphi_{,\mu},
\end{equation}
we obtain
\begin{equation}\label{eq:pr2q}
\Phi_{,\mu}^{,\mu}=2\varphi^{,\mu}\varphi_{,\mu}+2\varphi\varphi^{,\mu}_{,\mu}=
\frac{1}{2}\frac{\Phi^{,\mu}\Phi_{,\mu}}{\Phi}+2\varphi\varphi^{,\mu}_{,\mu}.
\end{equation}
Multiplying equation ~(\ref{eq:pr10}) by $2\varphi$, we obtain the following equations in terms of $\Phi$ instead of equations ~(\ref{eq:pr10}), (\ref{eq:pr11}), and (\ref{eq:pr12}):
\begin{equation}\label{eq:pr3q}
\Box\Phi-\frac{1}{2}\frac{\Phi^{,\mu}\Phi_{,\mu}}{\Phi}-2(e^2 B^\mu B_\mu-m^2)\Phi=0,
\end{equation}
\begin{equation}\label{eq:pr4q}
\Box B_\mu-B^\nu_{,\nu\mu}=-2e^2 B_\mu\Phi.
\end{equation}
To prove that these equations describe independent evolution of the electromagnetic field $B^\mu$, it is sufficient to prove that if components $B^\mu$ of the potential and their first derivatives with respect to $x^0$ ($\dot{B}^\mu$) are known in the entire space at some time point $x^0=\rm{const}$  (that means that all spatial derivatives of these values are also known in the entire space at that time point), equations ~(\ref{eq:pr3q}) and (\ref{eq:pr4q}) yield the values of their second derivatives, $\ddot{B}^\mu$, for the same value of $x^0$. Indeed, $\Phi$ can be eliminated using equation ~(\ref{eq:pr4q}) for $\mu=0$, as this equation does not contain $\ddot{B}^\mu$ for this value of $\mu$:
\begin{eqnarray}\label{eq:pr5q}
\nonumber
\Phi=(-2e^2 B_0)^{-1}(\Box B_0-B^\nu_{,\nu 0})=\\
(-2e^2 B_0)^{-1}(B^{,i}_{0,i}-B^i_{,i 0})
\end{eqnarray}
(Greek indices in the Einstein sum convention run from $0$ to $3$, and Latin indices run from $1$ to $3$).
Then $\ddot{B}^i$ ($i=1,2,3$) can be determined by substitution of equation ~(\ref{eq:pr5q}) into equation ~(\ref{eq:pr4q}) for $\mu=1,2,3$:
\begin{equation}\label{eq:pr6q}
\ddot{B}_i=-B^{,j}_{i,j}+B^\nu_{,\nu i}+(B_0)^{-1} B_i(B^{,j}_{0,j}-B^j_{,j 0}).
\end{equation}
Thus, to complete the proof, we only need to find $\ddot{B}^0$. Conservation of current implies
\begin{equation}\label{eq:pr7q}
0=(B^\mu \Phi)_{,\mu}=B^\mu_{,\mu}\Phi+B^\mu\Phi_{,\mu}.
\end{equation}
This equation determines $\dot{\Phi}$, as spatial derivatives of $\Phi$ can be found from equation ~(\ref{eq:pr5q}). Differentiation of this equation with respect to $x^0$ yields
\begin{eqnarray}\label{eq:pr8q}
\nonumber
0=(\ddot{B}^0+\dot{B}^i_{,i})\Phi+(\dot{B}^0+B^i_{,i})\dot{\Phi}+\\
\dot{B}^0\dot{\Phi}+\dot{B}^i\Phi_{,i}+B^0\ddot{\Phi}+B^i\dot{\Phi}_{,i}.
\end{eqnarray}
After substitution of $\Phi$ from equation ~(\ref{eq:pr5q}), $\dot{\Phi}$ from equation ~(\ref{eq:pr7q}), and $\ddot{\Phi}$ from  equation ~(\ref{eq:pr3q}) into  equation ~(\ref{eq:pr8q}), the latter equation determines $\ddot{B^0}$ as a function of $B^\mu$, $\dot{B}^\mu$ and their spatial derivatives (again, spatial derivatives of $\Phi$ and $\dot{\Phi}$ can be found from the expressions for $\Phi$ and $\dot{\Phi}$ as functions of $B^\mu$ and $\dot{B}^\mu$). Thus, if $B^\mu$ and $\dot{B}^\mu$ are known in the entire space at a certain value of $x^0$, then $\ddot{B}^\mu$ can be calculated for the same $x^0$, so integration yields $B^\mu$ in the entire space-time. Therefore, we do have independent dynamics of electromagnetic field.

A reader made the following important comment on a preliminary version of this article: ``The equations of scalar electrodynamics include (taking into account the gauge freedom) five real functions (for the fields $A_\mu$ and complex $\varphi$). These obey second order differential equations and hence one would naively expect that 10 initial conditions would be required. On the other hand, it seems that one only needs $B_\mu$ and $\dot{B}_\mu$ at $t=t_0$ in order to evolve in time; hence 8 initial conditions. A similar situation appears in the spinor case. The author should clarify the initial condition counting.'' The author would like to emphasize that the systems of partial differential equations considered in this article are very specific, and their peculiarities influence the choice of initial conditions of the Cauchy problem. The systems include both equations of motion and constraints, such as the Maxwell equation with index $\mu=0$, so standard methods developed for systems with constraints should be used to determine the true initial conditions. Such analysis, while straightforward, is beyond the scope of this article. The author only proved that electromagnetic field evolves independently (if we know $B_\mu$ and $\dot{B}_\mu$ at $t=t_0$, we can calculate $\ddot{B}_\mu$ at $t=t_0$), but that does not mean that, for example, arbitrary values of $B_\mu$ and $\dot{B}_\mu$ at $t=t_0$ can be chosen for scalar electrodynamics, as such arbitrary choice can be incompatible with the constraints. Some peculiarities that influence the choice of initial conditions: 1) equations (\ref{eq:pr7}), (\ref{eq:pr8}), and (\ref{eq:pr9}) are not independent, as current conservation can be derived both from the Maxwell equations and from the Klein-Gordon equation; 2) equation (\ref{eq:pr8q}), derived from the Maxwell equations, defines $\ddot{\Phi}$, so a constraint can be derived from equation (\ref{eq:pr8q}) and equation (\ref{eq:pr3q}).

\section{Spinor electrodynamics}
\label{sec:2}
The equations of spinor electrodynamics are as follows:
\begin{equation}\label{eq:pr25}
(i\slashed{\partial}-\slashed{A})\psi=\psi,
\end{equation}
\begin{equation}\label{eq:pr26}
\Box A_\mu-A^\nu_{,\nu\mu}=e^2\bar{\psi}\gamma_\mu\psi,
\end{equation}
where, e.g., $\slashed{A}=A_\mu\gamma^\mu$ (the Feynman slash notation). For the sake of simplicity, a system of units is used where $\hbar=c=m=1$, and the electric charge $e$ is included in $A_\mu$ ($eA_\mu \rightarrow A_\mu$).
In the chiral representation of $\gamma$-matrices (Ref.~\cite{Itzykson})
\begin{equation}\label{eq:d1}
\gamma^0=\left( \begin{array}{cc}
0 & -I\\
-I & 0 \end{array} \right),\gamma^i=\left( \begin{array}{cc}
0 & \sigma^i \\
-\sigma^i & 0 \end{array} \right),
\end{equation}
where index $i$ runs from 1 to 3, and $\sigma^i$ are the Pauli matrices.
Let us apply the following ``generalized gauge transform'':
\begin{equation}\label{eq:dd1}
\psi=\exp(i\alpha)\varphi,
\end{equation}
\begin{equation}\label{eq:dd1a}
A_\mu=B_\mu-\alpha_{,\mu},
\end{equation}
where the new four-potential $B_\mu$ is complex (cf. Ref.~\cite{Mignani}), but produces the same electromagnetic fields as $A_\mu$), $\alpha=\alpha(x^\mu)=\beta+i\delta$, $\beta=\beta(x^\mu)$, $\delta=\delta(x^\mu)$, and $\beta$, $\delta$ are real. The imaginary part of the complex four-potential is a gradient of a certain function, so alternatively we can use this function instead of the imaginary components of the four-potential.

After the transform, the equations of spinor electrodynamics can be rewritten as follows:
\begin{equation}\label{eq:dd2}
(i\slashed{\partial}-\slashed{B})\varphi=\varphi,
\end{equation}
\begin{equation}\label{eq:dd3}
\Box B_\mu-B^\nu_{,\nu\mu}=\exp(-2\delta)e^2\bar{\varphi}\gamma_\mu\varphi.
\end{equation}
If $\psi$ and $\varphi$ have components
\begin{equation}\label{eq:dd4}
\varphi=\left( \begin{array}{c}
\varphi_1\\
\varphi_2\\
\varphi_3\\
\varphi_4\end{array}\right),
\psi=\left( \begin{array}{c}
\psi_1\\
\psi_2\\
\psi_3\\
\psi_4\end{array}\right),
\end{equation}
let us fix the ``gauge transform'' of equations (\ref{eq:dd1}) and (\ref{eq:dd1a}) somewhat arbitrarily by the following condition:
\begin{equation}\label{eq:dd5}
\varphi_1=\exp(-i\alpha)\psi_1=1.
\end{equation}
The Dirac equation (\ref{eq:dd2}) can be written in components as follows:
\begin{widetext}
\begin{eqnarray}\label{eq:dd6}
(B^0+B^3)\varphi_3+(B^1-i B^2)\varphi_4+i(\varphi_{3,3}-i\varphi_{4,2}+\varphi_{4,1}-\varphi_{3,0})=\varphi_1,
\end{eqnarray}
\begin{eqnarray}\label{eq:dd7}
(B^1+i B^2)\varphi_3+(B^0-B^3)\varphi_4-i(\varphi_{4,3}-i\varphi_{3,2}-\varphi_{3,1}+\varphi_{4,0})=\varphi_2,
\end{eqnarray}
\begin{eqnarray}\label{eq:dd8}
(B^0-B^3)\varphi_1-(B^1-i B^2)\varphi_2-i(\varphi_{1,3}-i\varphi_{2,2}+\varphi_{2,1}+\varphi_{1,0})=\varphi_3,
\end{eqnarray}
\begin{eqnarray}\label{eq:dd9}
-(B^1+i B^2)\varphi_1+(B^0+B^3)\varphi_2+i\varphi_{2,3}+\varphi_{1,2}-i(\varphi_{1,1}+\varphi_{2,0})=\varphi_4.
\end{eqnarray}
\end{widetext}
Equations (\ref{eq:dd8}) and (\ref{eq:dd9}) can be used to express components $\varphi_3,\varphi_4$ via $\varphi_1,\varphi_2$ and eliminate them from equations (\ref{eq:dd6}) and (\ref{eq:dd7}). The resulting equations for $\varphi_1$ and $\varphi_2$ are as follows:
\begin{widetext}
\begin{eqnarray}\label{eq:dd10}
\nonumber
-\varphi_{1,\mu}^{,\mu}+\varphi_2(-i B^1_{,3}-B^2_{,3}+B^0_{,2}+B^3_{,2}+
i(B^0_{,1}+B^3_{,1}+B^1_{,0})+B^2_{,0})+\\
+\varphi_1(-1+B^{\mu} B_{\mu}-i B^{\mu}_{,\mu}+i B^0_{,3}-B^1_{,2}+B^2_{,1}+i B^3_{,0})-2i B^{\mu}\varphi_{1,\mu}=0,
\end{eqnarray}
\begin{eqnarray}\label{eq:dd11}
\nonumber
-\varphi_{2,\mu}^{,\mu}+i\varphi_1( B^1_{,3}+i B^2_{,3}+i B^0_{,2}-i B^3_{,2}+
B^0_{,1}-B^3_{,1}+B^1_{,0}+i B^2_{,0})+\\
+\varphi_2(-1+B^{\mu} B_{\mu}-i( B^{\mu}_{,\mu}+B^0_{,3}+i B^1_{,2}-i B^2_{,1}+ B^3_{,0}))-2i B^{\mu}\varphi_{2,\mu}=0.
\end{eqnarray}
\end{widetext}

Equation (\ref{eq:dd10}) can be used to express $\varphi_2$ via $\varphi_1$:
\begin{eqnarray}\label{eq:dd12}
\varphi_2=-\left(i F^1+F^2\right)^{-1}\left(\Box'+i F^3\right)\varphi_1,
\end{eqnarray}
where $F^i=E^i+i H^i$, real electric field $E^i$ and magnetic field $H^i$ are defined by the standard formulas
\begin{eqnarray}\label{eq:dd13}
F^{\mu\nu}=B^{\nu,\mu}-B^{\mu,\nu}=\left( \begin{array}{cccc}
0 & -E^1 & -E^2 & -E^3\\
E^1 & 0 & -H^3 & H^2\\
E^2 & H^3 & 0 & -H^1\\
E^3 &-H^2 & H^1 & 0  \end{array} \right),
\end{eqnarray}
and the modified d'Alembertian $\Box'$ is defined as follows:
\begin{eqnarray}\label{eq:dd14}
\Box'=\partial^\mu\partial_\mu+2 i B^\mu\partial_\mu+i B^\mu_{,\mu}-B^\mu B_\mu+1.
\end{eqnarray}
Equation (\ref{eq:dd11}) can be rewritten as follows:
\begin{eqnarray}\label{eq:dd15}
-\left(\Box'-i F^3\right)\varphi_2-\left(i F^1-F^2\right)\varphi_1=0.
\end{eqnarray}
Substitution of $\varphi_2$ from equation (\ref{eq:dd12}) into equation (\ref{eq:dd11}) yields an equation of the fourth order for $\varphi_1$:
\begin{widetext}
\begin{eqnarray}\label{eq:dd16}
\left(\left(\Box'-i F^3\right)\left(i F^1+F^2\right)^{-1}\left(\Box'+i F^3\right)-i F^1+F^2\right)\varphi_1=0.
\end{eqnarray}
\end{widetext}

Application of the gauge condition of equation (\ref{eq:dd5}) to equations (\ref{eq:dd14}), (\ref{eq:dd12}), (\ref{eq:dd16}), and (\ref{eq:dd15}) yields the following equations:
\begin{eqnarray}\label{eq:dd17}
\Box'\varphi_1=i B^\mu_{,\mu}-B^\mu B_\mu+1,
\end{eqnarray}
\begin{eqnarray}\label{eq:dd18}
\varphi_2=-\left(i F^1+F^2\right)^{-1}\left(i B^\mu_{,\mu}-B^\mu B_\mu+1+i F^3\right),
\end{eqnarray}
\begin{widetext}
\begin{eqnarray}\label{eq:dd19}
\left(\Box'-i F^3\right)\left(i F^1+F^2\right)^{-1}\left(i B^\mu_{,\mu}-B^\mu B_\mu+1+i F^3\right)-i F^1+F^2=0,
\end{eqnarray}
\end{widetext}
\begin{eqnarray}\label{eq:dd20b}
-\left(\Box'-i F^3\right)\varphi_2-\left(i F^1-F^2\right)=0.
\end{eqnarray}
Obviously, equations (\ref{eq:dd5}), (\ref{eq:dd18}), (\ref{eq:dd8}), and (\ref{eq:dd9}) can be used to eliminate spinor $\varphi$ from the equations of spinor electrodynamics (\ref{eq:dd2}) and (\ref{eq:dd3}). It is then possible to eliminate $\delta$ from the resulting equations. Furthermore, it turns out that the equations describe independent dynamics of the (complex four-potential of) electromagnetic field $B^\mu$. More precisely, if components $B^\mu$ and their temporal derivatives (derivatives with respect to $x^0$) up to the second order $\dot{B}^\mu$ and $\ddot{B}^\mu$ are known at some point in time in the entire 3D space $x^0$=const, the equations determine the temporal derivatives of the third order $\dddot{B}^\mu$, so the Cauchy problem can be posed, and the equations can be integrated (at least locally). Let us prove this statement.

As $\varphi_1$=1 (equation (\ref{eq:dd5})), we obtain
\begin{eqnarray}\label{eq:dd20}
\nonumber
\bar{\varphi}\gamma_\mu\varphi=\left( \begin{array}{c}
\varphi_1^*\varphi_1+\varphi_2^*\varphi_2+\varphi_3^*\varphi_3+\varphi_4^*\varphi_4\\
\varphi_2^*\varphi_1+\varphi_1^*\varphi_2-\varphi_4^*\varphi_3-\varphi_3^*\varphi_4\\
i\varphi_2^*\varphi_1-i\varphi_1^*\varphi_2-i\varphi_4^*\varphi_3+i\varphi_3^*\varphi_4\\
\varphi_1^*\varphi_1-\varphi_2^*\varphi_2-\varphi_3^*\varphi_3+\varphi_4^*\varphi_4\end{array}\right)=\\
\left( \begin{array}{c}
1+\varphi_2^*\varphi_2+\varphi_3^*\varphi_3+\varphi_4^*\varphi_4\\
\varphi_2^*+\varphi_2-\varphi_4^*\varphi_3-\varphi_3^*\varphi_4\\
i\varphi_2^*-i\varphi_2-i\varphi_4^*\varphi_3+i\varphi_3^*\varphi_4\\
1-\varphi_2^*\varphi_2-\varphi_3^*\varphi_3+\varphi_4^*\varphi_4\end{array}\right).
\end{eqnarray}
Using equation (\ref{eq:dd3}) with index $\mu=0$ and equation (\ref{eq:dd20}), we can express $e^2\exp(-2\delta)$ as follows:
\begin{eqnarray}\label{eq:dd21}
\nonumber
e^2\exp(-2\delta)=\\
\left(B^{,i}_{0,i}-B^i_{,i0}\right)
(1+\varphi_2^*\varphi_2+\varphi_3^*\varphi_3+\varphi_4^*\varphi_4)^{-1},
\end{eqnarray}
as
\begin{eqnarray}\label{eq:dd22a}
\Box B_0-B^\nu_{,\nu 0}=B^{,i}_{0,i}-B^i_{,i0}.
\end{eqnarray}
Substitution of equation (\ref{eq:dd21}) in equation (\ref{eq:dd3}) yields
\begin{eqnarray}\label{eq:dd23a}
\nonumber
\Box B_i-B^\nu_{,\nu i}=\ddot{B}_i+B_{i,j}^{,j}-\dot{B}^0_{,i}-B^j_{,j i}=\\
\nonumber
\left(B^{,j}_{0,j}-B^j_{,j0}\right)(1+\varphi_2^*\varphi_2+\varphi_3^*\varphi_3+\varphi_4^*\varphi_4)^{-1} \times \\
\left(\begin{array}{c}
\varphi_2^*+\varphi_2-\varphi_4^*\varphi_3-\varphi_3^*\varphi_4\\
i\varphi_2^*-i\varphi_2-i\varphi_4^*\varphi_3+i\varphi_3^*\varphi_4\\
1-\varphi_2^*\varphi_2-\varphi_3^*\varphi_3+\varphi_4^*\varphi_4\end{array}\right).
\end{eqnarray}
We are going to use equation (\ref{eq:dd23a}) first to express $\dddot{B}^i$ in terms of lower derivatives of $B^\mu$ with respect to time and then to express $\dddot{B}^0$ in terms of the said lower derivatives (see equation (\ref{eq:dd29}) below and the following equations of this Section).

We note based on equation (\ref{eq:dd18}) that $\varphi_2$ can be expressed via $B^\mu$, $\dot{B}^\mu$, and their spatial derivatives (derivatives with respect to $x^1$, $x^2$, and $x^3$), as
\begin{eqnarray}\label{eq:dd22}
\nonumber
F^1=E^1+i H^1=F^{10}+i F^{32}=\\
B^{0,1}-B^{1,0}+i(B^{2,3}-B^{3,2}),
\end{eqnarray}
\begin{eqnarray}\label{eq:dd22b}
\nonumber
F^2=E^2+i H^2=F^{20}+i F^{13}=\\
B^{0,2}-B^{2,0}+i(B^{3,1}-B^{1,3}),
\end{eqnarray}
\begin{eqnarray}\label{eq:dd22c}
\nonumber
F^3=E^3+i H^3=F^{30}+i F^{21}=\\
B^{0,3}-B^{3,0}+i(B^{1,2}-B^{2,1}).
\end{eqnarray}
Using equations (\ref{eq:dd22}), (\ref{eq:dd22b}), and (\ref{eq:dd22c}), the first temporal derivatives of $F^i$ can be written as follows:
\begin{equation}\label{eq:dd23}
\dot{F}^1=\dot{B}^{0,1}-\ddot{B}^{1}+i(\dot{B}^{2,3}-\dot{B}^{3,2}),
\end{equation}
\begin{equation}\label{eq:dd23b}
\dot{F}^2=\dot{B}^{0,2}-\ddot{B}^{2}+i(\dot{B}^{3,1}-\dot{B}^{1,3}),
\end{equation}
\begin{equation}\label{eq:dd23c}
\dot{F}^3=\dot{B}^{0,3}-\ddot{B}^{3}+i(\dot{B}^{1,2}-\dot{B}^{2,1}).
\end{equation}
We note based on equations (\ref{eq:dd18}), (\ref{eq:dd22}), (\ref{eq:dd22b}), (\ref{eq:dd22c}), (\ref{eq:dd23}), (\ref{eq:dd23b}), and (\ref{eq:dd23c}) that $\dot{\varphi}_2$ can be expressed via $B^\mu$, $\dot{B}^\mu$, $\ddot{B}^\mu$, and their spatial derivatives.

From equations (\ref{eq:dd8}) and (\ref{eq:dd5}) we obtain:
\begin{eqnarray}\label{eq:dd24}
\varphi_3=B^0-B^3-(B^1-i B^2)\varphi_2-i(-i\varphi_{2,2}+\varphi_{2,1}).
\end{eqnarray}
We note that $\varphi_3$ can be expressed via $B^\mu$, $\dot{B}^\mu$, and their spatial derivatives.
The first temporal derivative of $\varphi_3$ can be written as follows:
\begin{eqnarray}\label{eq:dd25}
\nonumber
\dot{\varphi}_3=\dot{B}^0-\dot{B}^3-(\dot{B}^1-i \dot{B}^2)\varphi_2-\\
(B^1-i B^2)\dot{\varphi}_2-i(-i \dot{\varphi}_{2,2}+ \dot{\varphi}_{2,1}).
\end{eqnarray}
We note based on equation (\ref{eq:dd25}) that $\dot{\varphi}_3$ can be expressed via $B^\mu$, $\dot{B}^\mu$, $\ddot{B}^\mu$, and their spatial derivatives.

From equations (\ref{eq:dd9}) and (\ref{eq:dd5}) we obtain:
\begin{eqnarray}\label{eq:dd26a}
\varphi_4=-(B^1+i B^2)+(B^0+B^3)\varphi_2+i\varphi_{2,3}-i\varphi_{2,0}.
\end{eqnarray}
We note that $\varphi_4$ can be expressed via $B^\mu$, $\dot{B}^\mu$, $\ddot{B}^\mu$, and their spatial derivatives.
The first temporal derivative of $\varphi_4$ can be written as follows:
\begin{eqnarray}\label{eq:dd26}
\nonumber
\dot{\varphi}_4=-(\dot{B}^1+i \dot{B}^2)+(\dot{B}^0+\dot{B}^3)\varphi_2+\\
(B^0+B^3)\dot{\varphi_2}+i\dot{\varphi}_{2,3}-i\ddot{\varphi}_2.
\end{eqnarray}
All terms in the expression for $\dot{\varphi}_4$ with a possible exception of $-i\ddot{\varphi}_2$ can be expressed via $B^\mu$, $\dot{B}^\mu$, $\ddot{B}^\mu$, and their spatial derivatives. Let us consider the expression $\ddot{\varphi}_2$. Equations (\ref{eq:dd20b}) and (\ref{eq:dd14}) yield:
\begin{eqnarray}\label{eq:dd27}
\nonumber
0=-\left(\Box'-i F^3\right)\varphi_2-\left(i F^1-F^2\right)=\\
\nonumber
-\left(\partial^\mu\partial_\mu+2 i B^\mu\partial_\mu+i B^\mu_{,\mu}-B^\mu B_\mu+1-i F^3\right)\varphi_2-\\
\nonumber
\left(i F^1-F^2\right)=\\
\nonumber
-\left(\partial^0\partial_0+\partial^i\partial_i+2 i B^0\partial_0+2 i B^i\partial_i\right)\varphi_2+\\
\nonumber
\left(i B^\mu_{,\mu}-B^\mu B_\mu+1-i F^3\right)\varphi_2-
\left(i F^1-F^2\right)=\\
\nonumber
-\ddot{\varphi}_2-2 i B^0\dot{\varphi}_2-\left(\partial^i\partial_i+2 i B^i\partial_i\right)\varphi_2+\\
\left(i B^\mu_{,\mu}-B^\mu B_\mu+1-i F^3\right)\varphi_2-
\left(i F^1-F^2\right).\;
\end{eqnarray}
We note that $\ddot{\varphi}_2$ can be expressed via $B^\mu$, $\dot{B}^\mu$, $\ddot{B}^\mu$, and their spatial derivatives. Therefore, based on equation (\ref{eq:dd26}), the same is true for $\dot{\varphi}_4$. Furthermore, we can summarize that all functions $\varphi_\mu$ and $\dot{\varphi}_\mu$ can be expressed via $B^\mu$, $\dot{B}^\mu$, $\ddot{B}^\mu$, and their spatial derivatives. Obviously, the same is true for $\varphi_\mu^*$ and $\dot{\varphi}_\mu^*$.

Differentiating equation (\ref{eq:dd23a}) with respect to time ($x^0$), we conclude that functions $\dddot{B}^i$ can be expressed via $B^\mu$, $\dot{B}^\mu$, $\ddot{B}^\mu$, and their spatial derivatives, as the left-hand side of equation (\ref{eq:dd23a}) after the differentiation equals
\begin{eqnarray}\label{eq:dd28}
\dddot{B}_i+\dot{B}_{i,j}^{,j}-\ddot{B}^0_{,i}-\dot{B}^j_{,j i},
\end{eqnarray}
and the right-hand side of equation (\ref{eq:dd23a}) after the differentiation will be expressed via $B^\mu$, $\dot{B}^\mu$, $\ddot{B}^\mu$, $\varphi_\mu$, $\dot{\varphi}_\mu$, $\varphi_\mu^*$, $\dot{\varphi}_\mu^*$, and their spatial derivatives. Therefore, functions $\dddot{B}_i$ can be expressed via $B^\mu$, $\dot{B}^\mu$, $\ddot{B}^\mu$, and their spatial derivatives, so to prove the initial statement we just need to prove the same for $\dddot{B}_0$. To this end, let us consider the following equation derived from equations (\ref{eq:dd19}) and (\ref{eq:dd14}):
\begin{widetext}
\begin{eqnarray}\label{eq:dd29}
\left(\partial^\mu\partial_\mu+2 i B^\mu\partial_\mu+i B^\mu_{,\mu}-B^\mu B_\mu+1-i F^3\right)
\left(i
F^1+F^2\right)^{-1}\left(i B^\mu_{,\mu}-B^\mu B_\mu+1+i F^3\right)-
i F^1+F^2=0.
\end{eqnarray}
\end{widetext}
It is obvious that the following part of the left-hand side of equation (\ref{eq:dd29}) can be expressed via $B^\mu$, $\dot{B}^\mu$, and their spatial derivatives:
\begin{eqnarray}\label{eq:dd30}
\nonumber
\left(i B^\mu_{,\mu}-B^\mu B_\mu+1-i F^3\right)\left(i
F^1+F^2\right)^{-1}\times\\
\left(i B^\mu_{,\mu}-B^\mu B_\mu+1+i F^3\right)-
i F^1+F^2.
\end{eqnarray}
The rest of the left-hand side of equation (\ref{eq:dd29}) can be rewritten as follows:
\begin{eqnarray}\label{eq:dd31}
\nonumber
\left(\partial^0\partial_0+\partial^i\partial_i+2 i B^0\partial_0+2 i B^i\partial_i\right)\left(i
F^1+F^2\right)^{-1}\times\\
\left(i B^\mu_{,\mu}-B^\mu B_\mu+1+i F^3\right).
\end{eqnarray}
The following part of the expression in equation (\ref{eq:dd31}) can be expressed via $B^\mu$, $\dot{B}^\mu$, and their spatial derivatives:
\begin{eqnarray}\label{eq:dd31a}
\nonumber
\left(\partial^i\partial_i+2 i B^i\partial_i\right)\left(i
F^1+F^2\right)^{-1}\times\\
\left(i B^\mu_{,\mu}-B^\mu B_\mu+1+i F^3\right).
\end{eqnarray}
Let us evaluate the following expression (as an intermediate step to evaluate the terms with $\partial^0\partial_0$ and $2 i B^0\partial_0$ in equation (\ref{eq:dd31})):
\begin{widetext}
\begin{eqnarray}\label{eq:dd32}
\nonumber
\partial_0\left(i F^1+F^2\right)^{-1}\left(i B^\mu_{,\mu}-B^\mu B_\mu+1+i F^3\right)=
-\left(i \dot{F}^1+\dot{F}^2\right)\left(i F^1+F^2\right)^{-2}\left(i B^\mu_{,\mu}-B^\mu B_\mu+1+i F^3\right)+\\
\left(i F^1+F^2\right)^{-1}\left(i \dot{B}^\mu_{,\mu}-2\dot{B}^\mu B_\mu+i \dot{F}^3\right).
\end{eqnarray}
\end{widetext}
It follows from equation (\ref{eq:dd32}) that the following part of equation (\ref{eq:dd31}) can be expressed via $B^\mu$, $\dot{B}^\mu$, $\ddot{B}^\mu$, and their spatial derivatives:
\begin{eqnarray}\label{eq:dd31ps}
\nonumber
\left(2 i B^0\partial_0\right)\left(i
F^1+F^2\right)^{-1}\left(i B^\mu_{,\mu}-B^\mu B_\mu+1+i F^3\right).
\end{eqnarray}

Therefore, we only need to evaluate (using equation (\ref{eq:dd32})) the following remaining part of equation (\ref{eq:dd31}):
\begin{widetext}
\begin{eqnarray}\label{eq:dd33}
\nonumber
\partial^0\partial_0\left(i F^1+F^2\right)^{-1}\left(i B^\mu_{,\mu}-B^\mu B_\mu+1+i F^3\right)=
\partial^0\left(-\left(i \dot{F}^1+\dot{F}^2\right)\left(i F^1+F^2\right)^{-2}\left(i B^\mu_{,\mu}-B^\mu B_\mu+1+i F^3\right)\right)+\\
\nonumber
\partial^0\left(i F^1+F^2\right)^{-1}\left(i \dot{B}^\mu_{,\mu}-2\dot{B}^\mu B_\mu+i \dot{F}^3\right)=
\partial^0\left(-\left(i \dot{F}^1+\dot{F}^2\right)\left(i F^1+F^2\right)^{-2}\left(i B^\mu_{,\mu}-B^\mu B_\mu+1+i F^3\right)\right)+\\
\left(\partial^0\left(i F^1+F^2\right)^{-1}\right)\left(i \dot{B}^\mu_{,\mu}-2\dot{B}^\mu B_\mu+i \dot{F}^3\right)+
\left(i F^1+F^2\right)^{-1}\left(i \dddot{B}^0+i \ddot{B}^i_{,i}+
\left(\partial^0\left(-2\dot{B}^\mu B_\mu+i \dot{F}^3\right)\right)\right).
\end{eqnarray}
\end{widetext}
It follows from equations (\ref{eq:dd23}), (\ref{eq:dd23b}), and (\ref{eq:dd23c}) that $\ddot{F^i}$ can be expressed via $B^\mu$, $\dot{B}^\mu$, $\ddot{B}^\mu$, $\dddot{B}^i$ (but not $\dddot{B}^0$), and their spatial derivatives, but, as explained above (see equation (\ref{eq:dd28}) and the text around it), $\dddot{B}^i$ can be expressed via $B^\mu$, $\dot{B}^\mu$, $\ddot{B}^\mu$, and their spatial derivatives. Thus, this is also true for all terms of the right-hand side (following the last equality sign) of equation (\ref{eq:dd33}) (and, consequently, for all terms of equation (\ref{eq:dd29})), with a possible exception of the term
\begin{eqnarray}\label{eq:dd33a}
\left(i F^1+F^2\right)^{-1}\dddot{B^0},
\end{eqnarray}
but that means that equation (\ref{eq:dd29}) can be used to express this term and, therefore, $\dddot{B^0}$ via $B^\mu$, $\dot{B}^\mu$, $\ddot{B}^\mu$, and their spatial derivatives, which completes the proof.

Thus, matter field can be eliminated from equations of scalar electrodynamics and spinor electrodynamics, and the resulting equations describe independent evolution of electromagnetic field (see precise wording above). It should be noted that these mathematical results allow different interpretations. For example, in the de Broglie - Bohm interpretation, electromagnetic field, rather than the wave function, can play the role of the guiding field ~\cite{Akhmeteli-IJQI}. Alternatively, one can also use the above results to get rid of the matter field altogether, as if it were just a ghost field, and leave just electromagnetic field in an interpretation. It seems that there may exist yet another interpretation of real charged fields shown in Refs.~\cite{Schroed,Akhmeteli-JMP} to be generally equivalent to complex fields in the context of the Klein-Gordon equation and the Dirac equation: the one-particle wave function may describe a large (infinite?) number of particles moving along the trajectories defined in the de Broglie - Bohm interpretation. The total charge, calculated as an integral of charge density over the infinite 3-volume, may still equal the charge of electron. So the individual particles can be either electrons or positrons, but all together they can be regarded as one electron, as the total charge is conserved.(If an electron is then removed, for example, as a result of a measurement, and the total energy of what is left is not very high, so it is difficult to speak about presence of pairs, then the remaining field will look very much like electronic vacuum, maybe with some electromagnetic  field.) This seems to be compatible with the notions of polarization of vacuum and path integral. So the wave function in a point can be a measure of polarization of vacuum in the point (and this may explain the fact that it determines the density of probability of finding a particle in this point), and spreading of wave packets should not create problems. This interpretation also seems to give a clearer picture of the two-slit interference. The author is not sure if such an interpretation has been proposed for ordinary complex charged fields, but it seems especially appropriate for real charged fields.

This interpretation may need a modification for composite particles, such as nucleons or large molecules, which also demonstrate quantum properties: it is difficult to imagine that molecule-antimolecule pairs play a significant role in diffraction of large molecules (creation of such pairs is possible, but much less probable than creation of electron-positron pairs). However, composite particles take part in some interaction (for example, electromagnetic or strong interactions), so the interpretation can be modified as follows in that case: composite particles are accompanied by a large collection of fermion-antifermion pairs (for example, electron-positron pairs for electromagnetic interactions and quark-antiquark pairs for strong interactions; in some situations, it can be difficult to tell such pairs from force carriers, such as photons or gluons). Such fermions are present at all locations where the wave function traditionally describing the composite particle does not vanish, so the dimensions of the collection are not limited by the range of the interaction (for example, the short range of strong interaction). Thus, the composite particle can be detected at all locations where the wave function does not vanish, although at most locations it is fermions of the collection that interact directly with the instrument, not the composite particle itself.

\section{Transition to Many-Particle Theories}
\label{sec:3}
To make this article more self-contained, this section contains a summary of the approach used in Ref.~\cite{Akhmeteli-IJQI} (following nightlight (Ref.~\cite{nigh})) to embed the resulting (non-second-quantized) theories describing independent evolution of electromagnetic field into quantum field theories. The following off-the-shelf mathematical result (Refs.~\cite{Kowalski,Kowalski2}), a generalization of the Carleman linearization, generates for a system of nonlinear partial differential equations a system of linear equations in the Fock space, which looks like a second-quantized theory and is equivalent to the original nonlinear system on the set of solutions of the latter.

Let us consider a nonlinear differential equation in an (s+1)-dimensional space-time (the equations describing independent dynamics of electromagnetic field for scalar electrodynamics and spinor electrodynamics are a special case of this equation) ${\partial_t}\boldsymbol{\xi}(x,t) = \boldsymbol{F}(\boldsymbol{\xi},{D^\alpha}\boldsymbol{\xi};x,t)$, $\boldsymbol{\xi}(x,0)=\boldsymbol{\xi}_0(x)$, where $\boldsymbol{\xi}:\mathbf{R}^s\times\mathbf{R}\rightarrow\mathbf{C}^k$ (function $\boldsymbol{\xi}$ is defined in an (s+1)-dimensional space-time and takes values in a $k$-dimensional complex space; for example, for spinor electrodynamics, the space-time is (3+1)-dimensional, and $\boldsymbol{\xi}$ includes real and imaginary parts of $B^\mu$, $\dot{B}^\mu$, and $\ddot{B}^\mu$), \linebreak $D^\alpha\boldsymbol{\xi}=\left(D^{\alpha_1}\xi_1,\ldots ,D^{\alpha_k}\xi_k\right)$, $\alpha_i$ are multiindices, ${D^\beta}={\partial^{|\beta|}}/\partial x_1^{\beta_1}\ldots\partial x_s^{\beta_s}$, with $ |\beta|=\sum\limits_{i=1}^{s}\beta_i$, is a generalized derivative, $\boldsymbol{F}$ is analytic in $\boldsymbol{\xi}$, $D^\alpha\boldsymbol{\xi}$. It is also assumed that $\boldsymbol{\xi_0}$ and $\boldsymbol{\xi}$ are square integrable. Then Bose operators $\boldsymbol{a^\dag(x)}=\left(a^\dag_1(x),\ldots,a^\dag_k(x)\right)$ and $\boldsymbol{a(x)}=\left(a_1(x),\ldots,a_k(x)\right)$ are introduced with the canonical commutation relations:
\begin{eqnarray}\label{eq:ad1}
\nonumber
\left[a_i(x),a^\dag_j(x')\right]=\delta_{ij}\delta(x-x')I,\\
\left[a_i(x),a_j(x')\right]=\left[a^\dag_i(x),a^\dag_j(x')\right]=0,
\end{eqnarray}
where $x,x'\in\mathbf{R}^s$, $i,j=1,\ldots,k$. Normalized functional coherent states in the Fock space are defined as $|\boldsymbol{\xi}\rangle =\exp\left(-\frac{1}{2}\int d^sx|\boldsymbol{\xi}|^2\right)\exp\left(\int d^sx\boldsymbol{\xi}(x)\cdot\boldsymbol{a}^\dagger(x)\right)|\boldsymbol{0}\rangle$. They have the following property:
\begin{equation}\label{eq:ad1a}
\boldsymbol{a}(x)|\boldsymbol{\xi}\rangle =\boldsymbol{\xi}(x)|\boldsymbol{\xi}\rangle.
\end{equation}
Then the following vectors in the Fock space can be  introduced:
\begin{widetext}
\begin{eqnarray}\label{eq:ad2}
|\xi,t\rangle = \exp\left[\frac{1}{2}\left(\int {d^s}x|\boldsymbol{\xi}|^2-\int {d^s}x|\boldsymbol{\xi}_0|^2\right)\right]|\boldsymbol{\xi}\rangle
=\exp\left(-\frac{1}{2}\int d^sx|\boldsymbol{\xi}_0|^2\right)
\exp\left(\int d^sx\boldsymbol{\xi}(x)\cdot\boldsymbol{a}^\dagger(x)\right)|\boldsymbol{0}\rangle.
\end{eqnarray}
\end{widetext}
Differentiation of equation ~(\ref{eq:ad2}) with respect to time $t$ yields, together with equation ~(\ref{eq:ad1a}), a linear Schr\"{o}dinger-like evolution equation in the Fock space:
\begin{eqnarray}\label{eq:ad3}
\frac{d}{dt}|\xi,t\rangle = M(t)|\xi,t\rangle,
|\xi,0\rangle=|\boldsymbol{\xi}_0\rangle,
\end{eqnarray}
where the boson ``Hamiltonian'' \linebreak
$M(t) = \int {d^s}x{\boldsymbol{a}^\dagger}(x)\cdot F(\boldsymbol{a}(x),{D^\alpha}\boldsymbol{a}(x))$. Let us note that the states of equation ~(\ref{eq:ad2}) are in general multi-particle states.

Obviously, the majority of solutions of the linear equations in the Fock space have no predecessors among the solutions of the initial nonlinear equations in (3+1) - dimensional space-time, so the strict principle of superposition is abandoned; however, there is a ``weak'' (or approximate) principle of superposition. Indeed, let us start with two different states in the Fock space corresponding (via the above procedure) to two different initial fields in 3 dimensions $\boldsymbol{\xi}(t_0,x)$ and $\boldsymbol{\psi}(t_0,x)$ (so these states are not the most general states in the Fock space). We can build a ``weak superposition'' of these states as follows: we build the following initial field in 3D: $a \boldsymbol{\xi}+b \boldsymbol{\psi}$, where $a$ and $b$ are the coefficients of the required superposition. Then we can build (using the above procedure) the state in the Fock space corresponding to $a \boldsymbol{\xi}+b \boldsymbol{\psi}$. If $\boldsymbol{\xi}$ and $\boldsymbol{\psi}$ are relatively weak, only the vacuum state and a term linear in $\boldsymbol{\xi}$ and $\boldsymbol{\psi}$ will effectively survive in the expansion of the exponent for the coherent state. However, what we typically measure is the difference between the state and the vacuum state. So we have approximate superposition, at least at the initial moment. However, as electrodynamic interaction is rather weak (this is the basis of QED perturbation methods), nonlinearity of the evolution equations in 3D is rather weak. Such analysis motivates that this ``superposition'' will not differ much from the ``true'' superposition of the states in the Fock space. We leave detailed analysis for future work.

\section{Conclusion}
\label{sec:concl}
A complex four-potential of electromagnetic field is introduced in the equations of spinor electrodynamics (the Dirac-Maxwell electrodynamics). This complex four-potential generates the same electromagnetic fields as the initial real four-potential. After that, the spinor field can be algebraically eliminated from the equations of spinor electrodynamics, and the resulting equations describe independent evolution of electromagnetic field. These equations are then embedded into a quantum field theory. At this stage, the theory provides a simple, valuable,  and at least reasonably realistic model, as it is based on spinor electrodynamics, but it is not clear if this model or its modification can describe experimental results as well as standard quantum electrodynamics.

Let us discuss just the following possible modification (nightlight, private communication). The theory of this work can be naturally extended to include Barut's self-field electrodynamics (SFED) (Refs.~\cite{Barutnato},~\cite{Barutvh}). Barut starts with equations of spinor electrodynamics, but eliminates the electromagnetic field using the Maxwell equations and the Green's function with $i\varepsilon$ prescription. ``As has been shown already by Feynman,~\cite{Feynman} the elimination of the electromagnetic field with the use of the Feynman propagator is completely equivalent to the quantization of the electromagnetic field as long as there are no photons present in the initial and final states.'' (Ref~\cite{Birula}) As a result, the electromagnetic field is complex in SFED (Ref~\cite{Birula}). However, this complex electromagnetic field still satisfies the Maxwell equations, as it is obtained from the equations using a Green's function. Therefore, to extend the theory of this article to SFED, we just need to consider a broader set of complex electromagnetic four-potentials $B^\mu$, not just those that can be made real by the ``generalized gauge transform'' (equations (\ref{eq:dd1}) and (\ref{eq:dd1a})). If we wish to limit ourselves to solutions of SFED, we need to consider a certain subset of this broader set of $B^\mu$. The possibility of extension to SFED is important as SFED reproduces some results of standard quantum electrodynamics with high accuracy ($\alpha^4$ and better -- cf. Ref.~\cite{Barutnato} and other articles on SFED in the same collection).

However, we should address the following weak point of the modification proposed in the previous paragraph: if the electromagnetic four-potential is complex, in general, the Dirac equation does not conserve the current. On the other hand, the Maxwell equations still imply current conservation. Therefore, the resulting equations are incompatible. The following approach can be used to resolve this difficulty. It was proven in Ref.~\cite{Akhmeteli-JMP} that the Dirac equation is generally equivalent to one complex equation for just one component of the Dirac spinor. The latter equation is generally equivalent to a system of two real equations (equations (26) and (27) of Ref.~\cite{Akhmeteli-JMP}), and the first of these equations describes current conservation. If the electromagnetic four-potential is complex, this equation does not hold (in general). Therefore, we can discard this equation and replace the Dirac equation in the equations of spinor electrodynamics by the second equation of the above-mentioned system (equation (27) of Ref.~\cite{Akhmeteli-JMP}) -- cf. Ref.~\cite{Schroed}. The resulting new system of equations will be equivalent (in general) to the equations of spinor electrodynamics if the electromagnetic four-potential is real (as the Maxwell equations imply current conservation). If the electromagnetic four-potential is complex, current conservation still holds (due to the Maxwell equations), and the new system of equations is probably compatible, but this has not been proven yet. Similar efforts might be needed to make room for the Feynman propagator for the spinor field.

\section*{Acknowledgments}
The author is grateful to A.V. Gavrilin, A.Yu. Khrennikov, nightlight, H. Nikoli\textrm{$\mathrm{\acute{c}}$}, W. Struyve, R. Sverdlov, and H. Westman for their interest in this work and valuable remarks.

The author is also grateful to J. Noldus for useful discussions.

\vspace*{-5pt}   

\end{document}